\documentstyle[12pt]{article}
\begin{document}
\begin{center}
{\Large
{\bf Covariant Quantization with Extended BRST Symmetry}}

\vspace{0.5cm}
{\sc B. Geyer $^{a)}$, D.M. Gitman $^{b)}$ and P.M. Lavrov $^{c)}$}\\
\vspace{0.5cm}
{\it ${}^{a)}$ Center of Theoretical Sciences and \\
Institute of Theoretical Physics, Leipzig University,\\
Augustusplatz 10/11, 04109 Leipzig, Germany}\\
\vspace{.3cm}
{\it ${}^{b)}$ Instituto de Fisica, Universidade de S\~{a}o Paulo\\
Caixa Postal 66318-CEP, 05315-970-S\~{a}o Paulo, S.P., Brazil}\\
\vspace{.3cm}
{\it ${}^{c)}$ Tomsk State Pedagogical University, 634041 Tomsk,
Russia
}

\end{center}
\vspace{.3cm}
\centerline{\bf Abstract}
A short rewiev of covariant quantization methods based on
BRST-antiBRST symmetry is given.
In particular problems of correct definition of $Sp(2)$ symmetric
quantization scheme known as triplectic quantization are considered.

\vspace{1cm}
\section{Introduction}
%\vspace{1cm}

It is well-known that all the fundamental interactions
(electromagnetic, strong, weak and gravitational) can be described in
terms of gauge theories \cite{1}. The quantization of gauge theories
is one of the most essential means to investigate the quantum
properties of the fundamental forces. The formalisms of Hamiltonian
(or canonical) and Lagrangian (or covariant) quantization of gauge
theories present two different approaches to the quantum description
of
dynamical systems \cite{2}.

The Lagrangian quantization remains one of the most attractive
approaches to gauge theories quantization owing to its main
advantage -- the possibility of direct construction of the quantum
effective action -- allowing to avoid the usual long way of canonical
quantization with subsequent integration over momenta in the path
integral, producing as a result the vacuum expectation value of the
S-matrix in the presence of external sources.
During last years covariant quantization schemes for arbitrary
gauge theories are intensively developed. The so-called
$Sp(2)$-covariant \cite{3}, superfield \cite{4}, triplectic \cite{5}
quantization methods have been discovered.
In addition, a further extension to an $osp(1,2)$--symmetric
quantization \cite{GLM}, allowing also for a superfield formulation
\cite{GM}, has been found.
These approaches are based
on the principle of invariance under global BRST-antiBRST symmetry
\cite{2a,3a,10a,11a} and are connected with different
off-shell realizations of this invariance principle in construction
of
Green's functions.

The purpose of this paper is to introduce the reader into the
$Sp(2)$-covariant \cite{3} and triplectic \cite{5} quantization
methods
and to present a modifed scheme of triplectic quantization \cite{6}.

\section{Sp(2)-covariant quantization}

Let us consider the Sp(2)-quantization for general gauge theory
described
by the initial classical action $S_0(A)$ of fields $A^i$ with
Grassmann parities $\epsilon (A^i)=\epsilon_i$. We assume that
$S_0(A)$
is invariant under gauge transformations
$\delta A^i=R^i_{\alpha}(A)\xi^{\alpha}$,
\begin{eqnarray}
S_{0,i}(A)R^i_{\alpha}(A)=0,
\end{eqnarray}
where $\xi^{\alpha}$ are arbitrary functions, and
$R^i_{\alpha}(A)$ are generators of gauge transformations. We have
also
used DeWitt's condensed notations \cite{DeW}, where any index
includes all
particular ones (space - time, index of internal group, Lorentz index
and
so on). Summation over repeated indices implies integration over
continuous
ones and usual summation over discrete ones.
We assume that the set $\{R^i_{\alpha}(A)\}$ is complete.
As a consequence of the condition of completeness, one can
prove \cite{7} that the algebra of generators has the following
general
form,
\begin{eqnarray}
%\label{GAGGT}
R^i_{\alpha ,
j}(A)R^j_{\beta}(A)-(-1)^{\varepsilon_{\alpha}\varepsilon_
{\beta}}R^i_{\beta ,j}(A)R^j_{\alpha}(A)=-R^i_{\gamma}(A)
F^{\gamma}_{\alpha\beta}(A)- S_{0,j}(A)M^{ij}_{\alpha\beta}(A),
\end{eqnarray}
where $F^{\gamma}_{\alpha\beta}(A)$ are structure functions
depending, in general, on the fields $A^i$ with the
following properties of symmetry $F^{\gamma}_{\alpha\beta}(A)=
-(-1)^{\varepsilon_{\alpha}\varepsilon_{\beta}}
F^{\gamma}_{\beta\alpha} (A)$
and $M^{ij}_{\alpha\beta}(A)$ satisfies the conditions
$M^{ij}_{\alpha\beta}(A) = -(-1)^{\varepsilon_i\varepsilon_j}
M^{ji}_{\alpha\beta}(A) =
-(-1)^{\varepsilon_{\alpha}\varepsilon_{\beta}}M^{ij}_{\beta\alpha}(A
)$.

Then it is necessary to introduce the total configuration space
$\phi^A$,
\begin{eqnarray}
%\label{ConfSpaceSp}
\phi^A = (A^i,\;B^{\alpha},\;C^{\alpha a},\cdot\cdot\cdot),\;\;\;
\varepsilon (\phi^A) = \varepsilon_A.
\end{eqnarray}
Here, $C^{\alpha a}$ are the $Sp(2)$--doublet of ghost ($a =1$) and
antighost ($a = 2$) fields  with respect to the index $a$,
$B^{\alpha}$ are auxiliary fields and dots denote (for reducible
theories) pyramids of ghosts, antighosts and Lagrange multipliers
which
are combined into irreducible representations of the symplectic
group $Sp(2)$ (see \cite{3}). Note that the general
ingredients and formulas have the same form both for irreducible and
reducible cases.

To each field $\phi^A$ of the total
configuration space one introduces three sets of antifields
$\phi^*_{Aa},\;\varepsilon(\phi^*_{Aa})=\varepsilon_A+1$
and $\bar\phi_A,\;\varepsilon(\bar\phi_A)=\varepsilon_A$,
being sources of extended BRST transformations,
namely, BRST-transformations, antiBRST-transformations and mixed
transformations respectively.

On the space of fields $\phi^A$ and antifields $\phi^*_{Aa}$ one
defines \cite{3} odd symplectic structures $(\;,\;)^a$, called the
extended antibrackets
\begin{eqnarray}
\label{AntiBSp}
(F,G)^a\equiv\frac{\delta F}{\delta\phi^A}\;\frac{\delta
G}{\delta\phi^*_{Aa}}
-(F\leftrightarrow G)\;(-1)^{(\varepsilon(F)+1)(\varepsilon(G)+1)}.
\end{eqnarray}
The derivatives with respect to fields are understood as acting
from the right and those with respect to antifields, as acting from
the
left. There are the graded Jacobi identities for the extended
antibrackets:
\begin{eqnarray}
%\nonumber
\label{PrAntiBSp}
 {((F,G)^{\{a},H)^{b\}}(-1)^{(\varepsilon(F)+1)(\varepsilon(H)+1)}
+{\rm cycl.perm.} (F,G,H)}\equiv 0,
\end{eqnarray}
where curly brackets denote symmetrization with respect to the
indices
$a,b$ of the $Sp(2)$ group.

In addition the operators $V^a,\;\Delta^a$ are introduced
\begin{eqnarray}
\label{VaSp}
V^a&=&\varepsilon ^{ab}\;\phi^*_{Ab}\;
\frac{\delta}{\delta\bar\phi_A},\\
\label{DeltaaSp}
\Delta^a&=&(-1)^{\varepsilon _A}\frac{\delta_{\it l}}{\delta\phi^A}
\frac{\delta}{\delta\phi^*_{Aa}},
\end{eqnarray}
where $\varepsilon ^{ab}$ is the antisymmetric tensor for raising and
lowering $Sp(2)$--indices
\begin{eqnarray}
\varepsilon ^{ab}=-\varepsilon ^{ba},\quad\varepsilon ^{12}=1\quad
\varepsilon_{ab}=-\varepsilon^{ab}.
\nonumber
\end{eqnarray}
It can be readily established that the algebra of the operators
(\ref{VaSp}), (\ref{DeltaaSp}) has the form
\begin{eqnarray}
\label{AlgebraDeltaaSp}
\Delta^{\{a}\Delta^{b\}}=0,\quad
\Delta^{\{a}V^{b\}}+
V^{\{a}\Delta^{b\}}=0,\quad
V^{\{a}V^{b\}}=0.
\end{eqnarray}
It is advantageous to introduce an operator
$\bar\Delta^a=\Delta^a+(i/\hbar)V^a $
with the properties
\begin{eqnarray}
\label{AlgBarDeltaSp}
\bar\Delta^{\{a}\bar\Delta^{b\}}=0.
\end{eqnarray}

For a boson functional $S=S(\phi,\phi^*,\bar\phi)$,
we introduce the extended quantum master equations
\begin{eqnarray}
\label{AcSp}
\frac{1}{2}(S,S)^a+V^aS=i\hbar\Delta^a S
\end{eqnarray}
with the boundary condition
\begin{eqnarray}
\label{}
S\big|_{\phi^* = \bar\phi = \hbar = 0} = S_0(A).
\end{eqnarray}

The generating equation for the bosonic functional $S$ is a set of
two
equations. It should be verified that these equations are compatible.
The simplest way to establish this fact is to rewrite the extended
master equations in the equivalent form of linear differential
equations
\begin{eqnarray}
\label{}
\bar\Delta^a \exp\left\{\frac{i}{\hbar}S\right\}=0.
\end{eqnarray}
Due to the properties (\ref{AlgBarDeltaSp}) of the operators
$\bar\Delta^a$,
we immediately establish the compatibility of the equations.

With the help of action (\ref{AcSp}) we next define the vacuum
functional $Z(0)$ by the rule
\begin{eqnarray}
\label{GFZSp}
 Z(0)&=&\int d\phi\;d\phi^{*}\;d\bar{\phi}\;d\lambda\;d\pi^a\;\exp
 \bigg\{\frac{i}{\hbar}\bigg(S(\phi,\phi^{*},\bar{\phi})+\phi^{*}_{Aa
}
 \pi^{Aa}+\nonumber\\&&
 +\bigg(\bar{\phi}_A-\frac{\delta F}
 {\delta\phi^A}\bigg)\lambda^A-\frac{1}{2}\varepsilon_{ab}\pi^{Aa}
 \frac{\delta^2F}{\delta\phi^A\delta\phi^B}\pi^{Bb}\bigg)\bigg\}\;,
\end{eqnarray}
where we have introduced a set of auxiliary fields
$\pi^{Aa},\varepsilon(\pi^{Aa})=\varepsilon_A+1,$,
$\lambda^A, \varepsilon(\lambda^A)=\varepsilon_A$
and a bosonic functional $F=F(\phi)$
fixing a gauge in the theory.

 An important property of the integrand in (\ref{GFZSp}) is its
invariance under the following global transformations (which, for its
 part, is a consequence of the extended master equation for $S$)
\begin{eqnarray}
\label{}
&& \delta\phi^A =\pi^{Aa}\mu_a,\quad
 \delta\phi^*_{Aa}=\mu_a\frac{\delta S}{\delta\phi^A},\quad
 \delta\bar{\phi}_A=\varepsilon^{ab}\mu_a\phi^*_{Ab},\nonumber\\
&& \delta\pi^{Aa}=-\varepsilon^{ab}\lambda^A\mu_b,\quad
 \delta\lambda^A=0,
\end{eqnarray}
 where $\mu_a$ is an Sp(2)--doublet of constant anticommuting
Grassmann
 parameters. These transformations realize the extended BRST
 transformations in the space of the variables $\phi$, $\phi^*$,
 $\bar{\phi}$, $\pi$ and $\lambda$.

 The existence of these transformations enables one to establish the
 independence of the vacuum functional from the choice of gauge.
 Indeed, suppose $Z_F\equiv Z(0)$. We shall change the gauge $F\to F+
 \Delta F$. In the functional integral for $Z_{F+\Delta F}$ we make
 the above-mentioned change of variables with the parameters chosen
as
\begin{eqnarray}
\label{}
 \mu_a=\frac{i}{2\hbar}\varepsilon_{ab}\frac{\delta\Delta
F}{\delta\phi^A}
 \pi^{Ab}.
\end{eqnarray}
 Then we find
\begin{eqnarray}
\label{}
 Z_F=Z_{F+\Delta F}
\end{eqnarray}
 and therefore the $S$-matrix is gauge-independent by virtue of the
equivalence theorem \cite{8}.

\section{Triplectic quantization}

The main idea of the triplectic quantization proposed by Batalin,
Marnelius and Semikhatov \cite{5} was to consider fields
$\pi^{Aa}$, which appear in the $Sp(2)$-method, as anticanonical
partners to the antifields $\bar{\phi}_A$ about extended antibrackets
defined by the rule:
\begin{eqnarray}
\label{AntiBTQ}
(F,G)^a\equiv\bigg(\frac{\delta F}{\delta\phi^A}\;\frac{\delta
G}{\delta\phi^*_{Aa}}+\epsilon^{ab}\frac{\delta F}{\delta\pi^{Ab}}\;
\frac{\delta G}{\delta\bar{\phi_A}}\;\bigg)
-(F\leftrightarrow G)\;(-1)^{(\varepsilon(F)+1)(\varepsilon(G)+1)}.
\end{eqnarray}
The new extended antibrackets have properties which formally coincide
with the properties of the extended antibrackets within the
$Sp(2)$--formalism
(see, for example, (\ref{PrAntiBSp})).
In an anloguous way the operators $V^a,\;\Delta^a$ are introduced by
\begin{eqnarray}
\label{VaTQ}
V^a=\frac{1}{2}\bigg(\varepsilon ^{ab}\;\phi^*_{Ab}
\frac{\delta}{\delta\bar{\phi}_A}-
(-1)^{\epsilon_A}\pi^{Aa}\frac{\delta_l}{\delta\phi^A}\bigg),
\end{eqnarray}
\begin{eqnarray}
\label{DeltaaTQ}
\Delta^a=(-1)^{\varepsilon _A}\frac{\delta_{\it l}}{\delta\phi^A}
\frac{\delta}{\delta\phi^*_{Aa}}+
(-1)^{\varepsilon _A+1}\epsilon^{ab}\frac{\delta_{\it
l}}{\delta\pi^{Aa}}
\frac{\delta}{\delta\bar {\phi}_A}.
\end{eqnarray}
It can be readily established that the algebra of operators
(\ref{VaTQ}), (\ref{DeltaaTQ}) has the form
\begin{eqnarray}
\label{AlVaTQ}
V^{\{a}V^{b\}}=0,\; \Delta^{\{a}\Delta^{b\}}=0,
\end{eqnarray}
\begin{eqnarray}
\label{AlDeltaaVaTQ}
\Delta^aV^b+V^b\Delta^a=0.
\end{eqnarray}
Note that the definition (\ref{VaTQ}) of the new operators $V^a$
differs from
the $Sp(2)$--case (see Eq.(\ref{VaSp})). As a consequence, formulas
(\ref{AlDeltaaVaTQ}) are valid without symmetrization
in the indices $a$ and $b$ in comparison with the $Sp(2)$--formalism.
It is also usefully to introduce an operator
$\bar\Delta^a=\Delta^a+(i/\hbar)V^a $
with the properties
\begin{eqnarray}
\label{AlbarDeltaaTQ}
\bar\Delta^{\{a}\bar\Delta^{b\}}=0.
\end{eqnarray}

The vacuum functional in this approach is defined by the rule
\begin{eqnarray}
\label{GFZTQ}
Z(0)=\int d\phi d\phi^* d\pi d\bar\phi
d\lambda\exp\bigg\{\frac{i}{\hbar}
(S+X)\bigg\}
\end{eqnarray}
where the boson functional $S=S(\phi,\phi^*,\pi,\bar\phi;\hbar)$
satisfies
the following master equations
\begin{eqnarray}
\label{MESeTQ}
\bar\Delta^a \exp\bigg\{\frac{i}{\hbar}S\bigg\}=0.
\end{eqnarray}
or, equivalently,
\begin{eqnarray}
\label{MESTQ}
\frac{1}{2}(S,\;S)^a+V^aS=i\hbar\Delta^a S,
\end{eqnarray}
and the boson functional
$X=X(\phi,\phi^*,\pi,\bar\phi,\lambda;\hbar)$
is a hypergauge fixing action depending on new variables
$\lambda^A,\;\epsilon (\lambda^A)=\epsilon_A$ and satisfing the
following
quantum equations:
\begin{eqnarray}
\label{MEGTQ}
\frac{1}{2}(X,\;X)^a-V^aX=i\hbar\Delta^a X,
\end{eqnarray}
which differs from Eq.(\ref{MESTQ}) by the opposite sign of the
V-term.

The vacuum functional (\ref{GFZTQ}) possesses the important property
of
invariance under the following global transformations
\begin{eqnarray}
\label{BRSTTQ}
\delta {\cal G} = ({\cal G},\;-S+X)^a\mu_a+2\mu_aV^a{\cal G},
\end{eqnarray}
 where ${\cal G}$ denotes the complete set of variables and
$\mu_a$ is an Sp(2) doublet of constant anticommuting Grassmann
 parameters. These transformations realize in the triplectic
quantization
 the extended BRST transformations in the space of the variables
 $\phi$, $\phi^*$, $\bar{\phi}$, $\pi^{a}$ and $\lambda$.

If we consider the transformations (\ref{BRSTTQ}) with $\mu_a$
depending on ${\cal G}$ and $\lambda$ it is not difficult
to obtain the following representation for the vacuum functional
\begin{eqnarray}
\label{VacFuncTQ}
Z(0)=\int d{\cal G}\exp\bigg\{\frac{i}{\hbar}
\bigg[S + X -i\hbar (\mu_a,S)^a + i\hbar (\mu_a,X)^a+
2i\hbar V^a\mu_a\bigg]\bigg\}
\end{eqnarray}
Let us make an additional change of variables in the integral
(\ref{VacFuncTQ})
\begin{eqnarray}
\label{Ghvar1TQ}
\delta {\cal G} = \frac{1}{2}({\cal G},\;\delta F_a)^a.
\end{eqnarray}
This change gives
\begin{eqnarray}
\label{}
Z(0)&=&\int d{\cal G}\exp\bigg\{\frac{i}{\hbar}
[S + X -i\hbar (\mu_a,S)^a + i\hbar (\mu_a,X)^a+
\nonumber\\
&&+ 2i\hbar V^a\mu_a+\frac{1}{2}(S,\;\delta F_a)^a +
\frac{1}{2}(X,\;\delta F_a)^a - i\hbar \Delta^a\delta F_a]\bigg\}
\end{eqnarray}
If we identify
\begin{eqnarray}
\label{}
\delta F_a({\cal G})\equiv\frac{2\hbar}{i}\mu_a({\cal G},\;\lambda)
\end{eqnarray}
then we obtain
\begin{eqnarray}
\label{}
Z(0)=\int d{\cal G}\exp\bigg\{\frac{i}{\hbar}
\bigg[S + X +\delta X\bigg]\bigg\}
\end{eqnarray}
where the following notation has been introduced
\begin{eqnarray}
\label{}
\delta X = (X,\;\delta F_a)^a - V^a\delta F_a -i\hbar\Delta^a\delta
F_a.
\end{eqnarray}
One can now check (for details, see \cite{5})
\begin{eqnarray}
\label{EqvarXTQ}
(X,\;\delta X)^a - V^a\delta X =i\hbar\Delta^a\delta X
\end{eqnarray}
provided $\delta F_a$ is chosen to have the following form
\begin{eqnarray}
\label{varFaTQ}
\delta F_a = \epsilon_{ab}\bigg\{(X,\;\delta Y)^b - V^b\delta Y
-i\hbar\Delta^b\delta Y\bigg\}.
\end{eqnarray}
On the other hand, any small admissible variation of hypergauge
fixing
action $\delta X$ in Eq.~(\ref{GFZTQ}) has to satisfy
Eqs.~(\ref{EqvarXTQ}). It means that one can compensate for a
variation of
hypergauge fixing action in the vacuum functional by a suitable
choice of
$\delta F_a$ in (\ref{Ghvar1TQ}) (or $\delta Y$ in (\ref{varFaTQ})).
Therefore, the vacuum functional (\ref{GFZTQ}) does not depend on the
gauge.
We see that from the formal point of view the triplectic quantization
\cite{5} possesses all remarkable features of the $Sp(2)$--method.

\section{Modified triplectic quantization}

Notice that in its original version \cite{5}, the classical action
$S_0(A)$ does not provide a solution of the master equations
(\ref{MESTQ}) because of the special structure of operators $V^a$
(\ref{VaTQ}) and cannot be considered as boundary condition to the
master equations. On the other hand  all known schemes of covariant
quantization and existence theorems for them are based on the fact
that all information of initial classical system is introduced
through
the boundary condition. This point should be considered as essential
part
of covariant quantizations.

We will show that it is possible to modify the original form of
triplectic
quantization \cite{5} preserving all attractive features of the
method
and allowing to encode in the usual way any information about the
classical
system with the help of the boundary condition to the master
equations.

We use the definition of extended antibrackets given in
(\ref{AntiBTQ}).
In solving the functional equations determining the effective action
we
make use of the operators $\Delta^a$, $V^a$ and $U^a$
\begin{eqnarray}
\label{DeltaaMTQ}
 \Delta^a&=&(-1)^{\varepsilon_A}\frac{\delta_l}{\delta\phi^A}
 \frac{\delta}{\delta\phi^{*}_{Aa}}
 +(-1)^{\varepsilon_A+1}\varepsilon^{ab}\frac{\delta_l}{\delta\pi^{Ab
}}
 \frac{\delta}{\delta\bar{\phi}_A},\\
\label{VaMTQ}
 V^a&=&\varepsilon^{ab}\phi^{*}_{Ab}\frac{\delta}{\delta\bar\phi_A}\;
,\\
\label{UaMTQ}
 U^a&=&(-1)^{\varepsilon_A+1}\pi^{Aa}\frac{\delta_l}{\delta\phi^A}\;.
\end{eqnarray}

Notice that the operators $\Delta^a$ have already appeared both
within the
scheme of triplectic quantization \cite{5} and, virtually, within the
scheme of superfield quantization \cite{4}. Even though the operators
$V^a$ in eq.~(\ref{VaMTQ}) differ from the corresponding operators of
the
triplectic quantization, %\cite{5}
 they coincide, at the same time, with
the operators applied in the framework of the $Sp(2)$ method
\cite{3}.
The use of the operators $U^a$ in eq.~(\ref{UaMTQ}) (an analog of
these
operators has been introduced in the method of superfield
quantization)
exhibits an essentially new feature as compared to both the
$Sp(2)$--method and the `old' triplectic quantization. % \cite{5}.

One readily establishes the following algebra of the operators
(\ref{DeltaaMTQ})--(\ref{UaMTQ}):
\begin{eqnarray}
\nonumber
&&\Delta^{\{a}\Delta^{b\}}=0,\quad V^{\{a}V^{b\}}=0,\quad
U^{\{a}U^{b\}}=0,\\
\label{Alg}
&&\Delta^{\{a}V^{b\}}+V^{\{a}\Delta^{b\}}=0,\quad
\Delta^{\{a}U^{b\}}+U^{\{a}\Delta^{b\}}=0,\\
\nonumber
&&V^aU^b+U^bV^a=0,\quad
\Delta^aV^b+V^b\Delta^a+\Delta^aU^b+U^b\Delta^a=0.
\end{eqnarray}
Apart from $\Delta^a$, $V^a$, $U^a$, we also introduce the operators
$\bar{\Delta}^a\equiv\Delta^a+(i/\hbar)V^a,\quad
 \tilde{\Delta}^a\equiv\Delta^a-(i/\hbar)U^a$.
With the above mentioned properties it follows that the
algebra of these operators has the form
\begin{eqnarray}
\bar{\Delta}^{\{a}\bar{\Delta}^{b\}}=0,\quad
\tilde{\Delta}^{\{a}\tilde{\Delta}^{b\}}=0,\quad
\bar{\Delta}^{\{a}\tilde{\Delta}^{b\}}+
 \tilde{\Delta}^{\{a}\bar{\Delta}^{b\}}=0.
\end{eqnarray}
%which follows from eqs. (\ref{Alg})

Let us denote by $S=S(\phi,\phi^*,\pi,\bar{\phi})$ the quantum
action,
corresponding to the initial classical theory with the action
$S_0=S_0(A)$, and defined as a solution of the following master
equations:
\begin{eqnarray}
\label{MEMTQ}
 \bar{\Delta}^a\exp\left\{\frac{i}{\hbar}S\right\}=0.
\end{eqnarray}
with the standard boundary condition
\begin{eqnarray}
 \left.S\right|_{\phi^*=\bar{\phi}=\hbar=0}=S_0.
\end{eqnarray}
Let us further define the vacuum functional as the following
functional
integral (${\cal G}=(\phi, \phi^*, \pi,\bar{\phi}, \lambda)$):
\begin{eqnarray}
\label{ZMTQ}
 Z=\int d{\cal G}\exp\left\{
 \frac{i}{\hbar}\left(S+X+\phi^*_{Aa}\pi^{Aa}\right)\right\},
\end{eqnarray}
where $X=X(\phi,\phi^*,\pi,\bar{\phi},\lambda)$ is a bosonic
functional
depending on the new variables $\lambda^A$,
$\varepsilon(\lambda)=\varepsilon_A$, which serve as gauge-fixing
parameters. We require that the functional $X$ satisfies the
following master
equation:
\begin{eqnarray}
\label{GEXMTQ}
 \tilde{\Delta}^a\exp\left\{\frac{i}{\hbar}X\right\}=0.
\end{eqnarray}
Notice that the generating equations determining the quantum action
$S$ in
eq.~(\ref{MEMTQ}) and the gauge-fixing functional $X$ in
eq.~(\ref{GEXMTQ}) differ -- along with the vacuum functional $Z$ in
eq.~(\ref{ZMTQ}) -- from the corresponding definitions applied in the
method of triplectic quantization \cite{5}.

One can readily obtain the simplest solution of eq.~(\ref{GEXMTQ})
determining the gauge-fixing functional $X$
\begin{eqnarray}
\label{XsMTQ}
 X&=&\left(\bar{\phi}_A-\frac{\delta
F}{\delta\phi^A}\right)\lambda^A-
 \frac{1}{2}\varepsilon_{ab}U^aU^bF=\nonumber\\
 &=&\left(\bar{\phi}_A-\frac{\delta F}{\delta\phi^A}\right)\lambda^A
 -\frac{1}{2}\varepsilon_{ab}\pi^{Aa}{\frac {\delta^2 F}{\delta\phi^A
 \delta\phi^B}}\pi^{Bb},
\end{eqnarray}
where $F=F(\phi)$ is a bosonic functional depending only on the
fields
$\phi^A$. As a straightforward exercise, one makes sure that the
functional $X$ in eq.~(\ref{XsMTQ}) does satisfy eq.~(\ref{GEXMTQ}).
If we
further  demand that the quantum action $S$ does not depend on the
fields $\pi^A$, then the functional (\ref{ZMTQ}) becomes exactly the
vacuum functional of the $Sp(2)$ quantization scheme \cite{3}.

Let us consider a number of properties inherent in the present scheme
of
triplectic quantization.
In the first place, the vacuum functional (\ref{ZMTQ}) is invariant
under the following transformations:
\begin{eqnarray}
\label{BRSTMTQ}
 \delta{\cal G}=({\cal G},-S+X)^a\mu_a+\mu_a(V^a+U^a){\cal G},
\end{eqnarray}
where $\mu_a$ is an $Sp(2)$ doublet of constant anticommuting
 parameters.  The transformations (\ref{BRSTMTQ}) play the role of
the
transformations of extended BRST symmetry, realized on the space of
the
variables ${\cal G}=(\phi$, $\phi^*$, $\pi$, $\bar{\phi}, \lambda)$.

Consider now the question of gauge dependence in the case of the
vacuum
functional $Z$ (\ref{ZMTQ}). Any admissible variation $\delta X$
should
satisfy the equations
\begin{eqnarray}
\label{VarXMTQ}
(X,\delta X)^a-U^a\delta
 X=i\hbar\Delta^a\delta X.
\end{eqnarray}
It is convenient to consider
an $Sp(2)$--doublet of operators $\hat{S}^a(X)$, defined by the rule
\begin{eqnarray}
 (X,F)^a\equiv\hat{S}^a(X)\cdot F,
\end{eqnarray}
and possessing the properties
\begin{eqnarray}
 \hat{S}^{\{a}(X)\hat{S}^{b\}}(X)=\hat{S}^{\{a}\left
 (\frac{1}{2}(X,X)^{b\}}\right),
\end{eqnarray}
which follow from the generalized Jacobi identities
(\ref{PrAntiBSp}).
Consequently,  eq.~(\ref{VarXMTQ}) can be represented in the form
\begin{eqnarray}
\label{a}
\hat{Q}^a(X)\delta X=0,
\end{eqnarray}
where we have
introduced an $Sp(2)$--doublet of nilpotent anticommuting operators
$\hat{Q}^a$, defined by the rule
$\hat{Q}^a(X)=\hat{S}^a(X)-i\hbar\tilde{\Delta}^a$,
 $\hat{Q}^{\{a}(X)\hat{Q}^{b\}}(X)=0$.
Then any functional of the form
\begin{eqnarray}
\label{b}
\delta X=\frac{1}{2}\varepsilon_{ab}\hat{Q}^a(X)\hat{Q}^b(X)\delta Y,
\end{eqnarray}
with an arbitrary bosonic functional $\delta Y$, is a solution of
eq.~(\ref{a}).  Moreover, by analogy with the theorems proved in
Ref.~\cite{3}, one establishes the fact that any solution of
eq.~(\ref{a}) -- vanishing when all the variables in $\delta X$ are
equal to
zero -- has the form (\ref{b}), with a certain bosonic functional
$\delta Y$.

Let us denote by $Z_X\equiv Z$ the value of the vacuum functional
(\ref{ZMTQ}) corresponding to the gauge condition chosen as a
functional $X$.
In the vacuum functional $Z_{X+\delta X}$ we first make the change of
variables (\ref{BRSTMTQ}), with $\mu_a=\mu_a({\cal G},\lambda)$, and
then, accompanying it with a subsequent change of variables
 $\delta\Gamma=({\cal G},\delta Y_a)^a,\;\;\;\varepsilon(\delta
Y_a)=1$,
with $\delta Y_a=-i\hbar\mu_a({\cal G},\lambda)$, we arrive at
\begin{eqnarray}
\label{VarZMTQ}
 Z_{X+\delta X}=\int d{\cal G}
 \exp\left\{\frac{i}{\hbar}\bigg(S+X+\delta X+\delta X_1
 +\Phi^*_{Aa}\pi^{Aa}\bigg)\right\}.
\end{eqnarray}
In eq.~(\ref{VarZMTQ}) we have used the notation
\begin{eqnarray}
 \delta X_1=2\bigg((X,\delta Y_a)^a-U^a\delta
Y_a-i\hbar\Delta^a\delta
 Y_a\bigg)=2\hat{Q}^a(X)\delta Y_a.
\end{eqnarray}
Let us choose the functional $\delta Y_a$ in the form
\begin{eqnarray}
 \delta Y_a=\frac{1}{4}\varepsilon_{ab}\hat{Q}^b\overline{\delta
Y},\;\;\;
 \varepsilon(\overline{\delta Y})=0.
\end{eqnarray}
Then, representing $\delta X$ as in eq.~(\ref{b}), and identifying
$\delta Y=-\overline{\delta Y}$, we find that
\begin{eqnarray}
 Z_{X+\delta X}=Z_X,
\end{eqnarray}
i.e.,~the vacuum functional (and also the $S$ matrix)
does not depend on the choice of gauge.

\section{Discussion}
The reader may profit by considering the original version
of triplectic quantization \cite{5} as compared to the modified
scheme,
proposed in \cite{6}. Thus, both versions are based on extended
BRST symmetry. Both versions apply the vacuum functional and the $S$
matrix not depending on the choice of gauge.
Both
versions implement the idea of separate treatment of the quantum
action
and the gauge-fixing functional, based each on appropriate master
equations. The principal distinctions concern a different form
of these equations as well as a different form of the vacuum
functional. The modification of the generating equations \cite{5}
permits incorporating the information contained in the initial
classical action by means of the corresponding boundary conditions.
In
contrast to the original version \cite{5}, the classical action
provides a solution of the modified master equation. Thus, one
establishes a connection with the previous schemes of covariant
quantization. In particular, one easily reveals the fact of
equivalence
with the $Sp(2)$ quantization, by means of explicit realization of
the
corresponding class of boundary condition. In the original version of
triplectic quantization, however, these questions still remained
open.

 Another distinction of the two triplectic quantization schemes is
connected with the explicit structure of the corresponding master
equations. Thus, the original version \cite{5} of triplectic
quantization defined the generating equations for the quantum action
and the vacuum functional, using the operators
 \begin{eqnarray}
\label{Vbm}
V_{\rm BM}^a=\frac{1}{2}\left(
 \varepsilon^{ab}\phi^{*}_{Ab}\frac{\delta}{\delta\bar\phi_A}
 +\pi^{Aa}(-1)^{\varepsilon_A+1}\frac{\delta_l}{\delta\phi^A}\right)
 =\frac{1}{2}(U^a+V^a).
\end{eqnarray}

The use of the generating equations determining the quantum action
with
the help of the operators $V_{\rm BM}^a$ leads to the following
characteristic feature of the triplectic quantization \cite{5}: the
classical action of the initial theory, defined as a limit of the
quantum
action at $\hbar\to 0$ and $\phi^*=\bar{\phi}=\pi=0$, does not
satisfy
the generating equations of the method. In turn, the proofs of the
existence theorems for the generating equations in all known methods
of Lagrangian quantization are based on the fact that the initial
classical action is a solution of the corresponding master equations.
Moreover, from the viewpoint of the superfield quantization \cite{4},
which applies operators $V^a$, $U^a$, whose component representation
is
\begin{eqnarray}
\label{c}
 V^a&=&\varepsilon^{ab}\phi^{*}_{Ab}\frac{\delta}{\delta\bar\phi_A}
 -J_A\frac{\delta}{\delta\Phi^{*}_{Aa}},\nonumber\\
 \\
 U^a&=&(-1)^{\varepsilon_A+1}\pi^{Aa}
 \frac{\delta_l}{\delta\phi^A}
 + (-1)^{\varepsilon_A}\varepsilon^{ab}\lambda^A
 \frac{\delta_l}{\delta\pi^{Ab}}\nonumber
\end{eqnarray}
(with $J_A$ being the sources to the fields $\phi^A$), the operators
(\ref{Vbm})
have no precise geometrical meaning, whereas the $V^a$ and $U^a$ in
eq.~(\ref{c}) serve as generators of supertranslations -- in
superspace
spanned by superfields and superantifields -- along additional
(Grassmann) coordinates. In turn, the operators $V^a$ (\ref{VaMTQ})
and
$U^a$ (\ref{UaMTQ}) can be considered as limits (at $J_A=0$,
$\lambda^A=0$) of the operators (\ref{c}), which possess a clear
geometrical meaning.

The present modified scheme of triplectic
quantization enjoys every attractive feature of the quantization
\cite{5}: the theory possesses extended BRST transformations; the
vacuum functional and the $S$ matrix do not depend on the choice of
the
gauge-fixing functional; there exists such a choice of the
gauge-fixing
functional and solutions of the generating equations that reproduces
the results of the $Sp(2)$ method.

\section*{Acknowledgments}
The work of one of the authors (P.M.L) has been supported by the
Russian
Foundation for Basic Research, project 99-02-16617, as well as by
grant INTAS 96-0308 and grant RFBR-DFG 96-02-00180.  D.M.G. thanks
Brazilian Foundation CNPq and FAPESP for
for partial support.

%%%%%%%%%%%%%%%%%%%%%%%%%%%

\end{document}